# Acoustofluidic Engineering of Functional Vessel-on-a-Chip

Yue Wu[1], Yuwen Zhao[1], Khayrul Islam[2], Yuyuan Zhou[1], Saeed Omidi[1], Yevgeny Berdichevsky[1, 3], and Yaling Liu[1, 2, *)]

## Affiliation

1. Department of Bioengineering, Lehigh University, Bethlehem, Pennsylvania 18015, USA
2. Department of Mechanical Engineering and Mechanics, Lehigh University, Bethlehem, Pennsylvania 18015, USA
3. Department of Electrical and Computer Engineering, Lehigh University, Bethlehem, Pennsylvania 18015, USA

*) Author to whom correspondence should be addressed: yal310@lehigh.edu

## ABSTRACT

Construction of *in vitro* vascular models is of great significance to various biomedical research, such as pharmacokinetics and hemodynamics, and thus is an important direction in tissue engineering field. In this work, a standing surface acoustic wave field was constructed to spatially arrange suspended endothelial cells into a designated acoustofluidic patterning. The cell patterning was maintained after the acoustic field was withdrawn within solidified hydrogel. Then, interstitial flow was provided to activate vessel tube formation. In this way, a functional vessel network with specific vessel geometry was engineered on-chip. Vascular function, including perfusability and vascular barrier function, was characterized by microbeads loading and dextran diffusion, respectively. A computational atomistic simulation model was proposed to illustrate how solutes cross vascular lipid bilayer. The reported acoustofluidic methodology is capable of facile and reproducible fabrication of the functional vessel network with specific geometry and high resolution. It is promising to facilitate the development of both fundamental research and regenerative therapy.

## INTRODUCTION

Vascular system is one of the most important circulatory systems in the human body[1]. Blood vessels are not only a necessary pathway for physiological metabolism, material exchange, and nutrient delivery, but also an important route for most pharmacokinetic drug delivery[2,3]. In order to match specific different structures and functions of each organ, the blood vessels have significant organ specificity[4,5]. For example, the vascular structure in the hepatic lobule has a hexagonal distribution[6], while the vessels in muscle tissue run in parallel lines[7]. Therefore, the *in vitro* reconstruction of functional vascular models has always been one of the key research topics in the field of tissue engineering[8].

At present, self-assembly nature of endothelial cells (ECs) is mostly used for *in vitro* capillary vessel network construction[9–13], with inevitable randomness and low reproducibility[14]. In an attempt to control the vessel shape, researchers have been building blood vessels by coating the inner surface of hollow gel channels with ECs[15–17]. However, limited by the template resolution and difficulty of removing mold template without destroying channel, it is difficult to achieve micron-scale blood vessels with slightly complex shapes[18,19]. Multiphoton ablation technology reported by Prof. Zheng *et al.* can generate hollow gel channels with intricate patterns of a few microns, thus accurately control the shape and size of blood vessels at the same time[20–22]. However, the complexity and cost of the laser machine system may limit its widespread adoption. Due to various strengths, such as remote control, biocompatibility, non-contact and easy operation, acoustofluidics has gradually developed into a popular biomedical technology[23,24], and widely used in the fields of disease diagnosis[25–30], drug delivery[31–34], tissue engineering[35–40] and biophysical characterization[41,42]. Here, we report a standing surface acoustic wave (SSAW)- based acoustofluidic methodology as an alternate of engineering vessel-on-a-chip. First, the suspended endothelial cells were acoustically patterned into parallel lines topography in the hydrogel matrix. Then, the patterned endothelial cells were activated by the interstitial flow and developed into functional vessel tubes along the previous acoustic patterning geometry. The permeability difference between the mono-vessel and fibroblast-supported vessel was evaluated and a computational model was proposed for interpretation. Last, by characterizing the compound response of neurons in the pure matrix and in the vessel-associated matrix, the influence of blood vessels was demonstrated. The reported acoustofluidic vessel engineering demonstrates that without any physical channel as a guide, the HUVECs can still assembled into the vessel network with the designated specific geometry. Besides, the cost of the acoustofluidic system is more lab-friendly than photon laser facility. Thus, it is promising to be an alternative method for facile and high-resolution construction of *in vitro* vessel model.

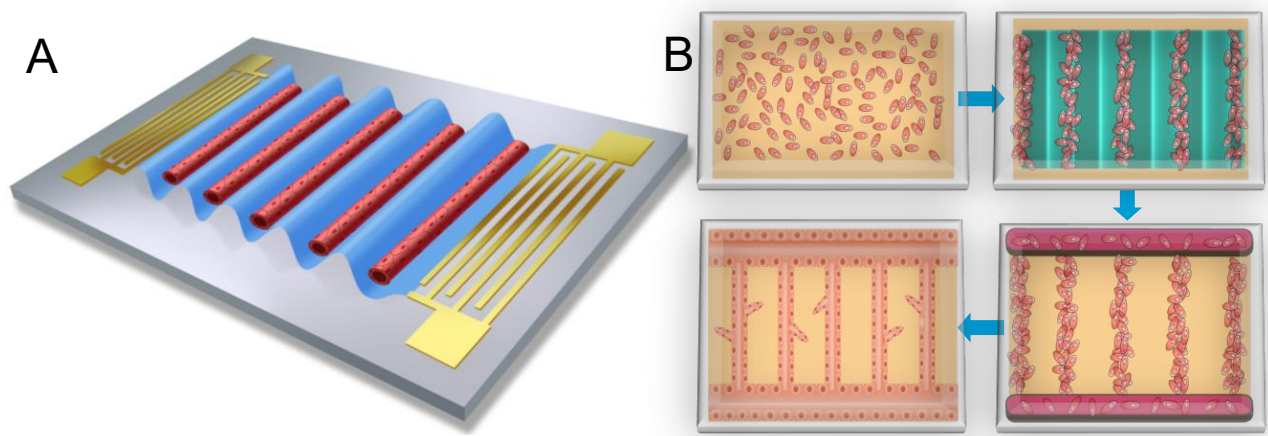

Figure 1. Schematics of acoustofluidic engineering functional vessel-on-a-chip. A) As shown schematically, standing surface acoustic wave is generated on-chip to pattern the cells and promote the desired vessel formation. B) Flow chart of the acoustofluidic engineering of vessel formation. The suspended cells are patterned into the acoustic pressure node and aligned into parrel line array. After the acoustic field disappears, the solidified gel can still maintain the original patterning shape of the cells. Due to the nature of adherent growth, some excess vascular cells will grow along the wall of the PDMS chamber, forming transverse vessels perpendicular to the patterned longitudinal vessel tubes, thereby connecting these parallel arrays of vascular tubes. In the end, functional vessel-on-chip can be formed with the interstitial flow stimulation.

## MATERIALS AND METHODS

### Acoustofluidic device fabrication

The SSAW device was fabricated through standard lithography and lift-off process[43]. A 15-μm-thick photoresist layer (AZ512, Kayaku Advanced Materials, Inc.) was spin-coated on a 500-μm-thick, double-side polished, piezoelectric $LiNbO_3$ wafer (Precision Micro-Optics, Inc.). Then, the designed interdigital transducers (IDT) of double metal layers (Cr/100 Å, Au/600 Å) with a 125 μm finger width were transferred from the customized plastic mask to the substrate by metal deposition[44]. The finger pair number of the IDT is 30. The resonant frequency of the obtained SSAW device was measured as around 8.21 MHz. The microfluidic chamber mold was fabricated with soft lithography[45]. A 150-μm-thick SU8-2150 (Kayaku Advanced Materials, Inc.) photoresist mold was created on a 3-inch silicon wafer. After being peeled off from the mold, the polydimethylsiloxane (PDMS) chambers were plasma-bonded on a 1 mm thick glass coverslip[46]. All the fabrication processes were finished in the Center for Photonics and Nanoelectronics (CPN) at Lehigh University.

### Cell culture, experimental setup and immune-fluorescence staining

Human Umbilical Vein Endothelial Cells (HUVEC) and Normal Human Lung Fibroblasts (NHLF) were purchased from LONZA[47]. HUVECs were cultured in an endothelial cell growth medium supplemented with EGM-2 SingleQuot kit supply and growth factors (EGM-2, LONZA). NHLF were cultured in fibroblast growth basal medium supplemented with FGM™-2 BulletKit™ (FGM-2, LONZA)[48]. The generation of excitatory human forebrain neurons was accomplished using over expression of Neurogenin-2 (NGN2) transcription factor, following the protocol[49,50]. The neurons were cultured with complete mTeSR1 medium, differentiation to neuron happened with introducing dox to the culture (N2 supplement + doxycycline and growth factors in DMEM/F12 media) and maintained in culture with 50x B27 supplement and 2 mM L-Glutamine in Neurobasal medium (Thermo Fisher Scientific). All cells were cultured at 37 °C in a humidified 5% $CO_2$ environment.

For the patterning experiment, bovine fibrinogen (CAS-9001-32-5, Sigma) was dissolved in Dulbecco's phosphate buffered saline (DPBS) to 2 mg/ml solution. The cell-laden fibrinogen pre gel solution was mixed with 2U/ml bovine thrombin (CAS-9002-044, Sigma) and injected into the chamber through chamber loading ports[51]. For the pure HUVEC condition, the cell concentration was 5 million/ml. For the coculture situation, the HUVEC concentration retained with 5 million/ml, while the concentration of another cell was 0.2 million/ml. During the cell patterning experiment, a radio frequency signal from a function generator (AFG3102, Tektronix, USA) was amplified by an amplifier (ACS-230-25W, Com-Power, USA), and transferred to the IDT pair. The input voltages on the devices were from 50 Vpp. A coupling water layer was placed between the SSAW device and the PDMS-glass-based microfluidic chamber for acoustic wave conduction[52,53]. The HUVECs were aligned into parallel lines in PDMS channels during 3 minutes of acoustic patterning process. Then the PDMS chambers were relocated into the incubator for the gel solidification and further *in vitro* cell culture[54].

For immunofluorescence staining of the vessel model, detailed method can be found in our previous publication[13]. For the neural cells-vessel coculture system, 1.0 mM monosodium glutamate (MSG) (Sigma-Aldrich) was dissolved in EGM medium and loaded into the vessel tube. c-Fos (sc-166940-AF488, Santa Cruz)[55] immunostaining was executed 10 min after loading the monosodium glutamate solution. All the fluorescent images and confocal scanning were acquired with a Nikon C2+ laser scanning confocal microscope in the Health Research Hub Center, Lehigh University. The image processing was conducted with ImageJ software from National Institutes of Health (USA).

**On-chip vessel network construction and vascular permeability characterization**

After the gel curing, a pipette tip containing 50 μL culture medium were inserted into one of the chamber loading ports to provide interstitial flow and hydrostatic, while an empty pipette tip was inserted into the other chamber loading port for medium collection[13,47]. The culture medium in pipette tips were refreshed every day. The permeability of the vessels was measured with solute diffusion across the vessel wall. 70kDa FITC conjugated dextran (CAS-60842-46-8, Sigma) was dissolved in the EGM-2 medium to prepare the solution[56,57].

**Computational simulation**

COMSOL® Multiphysics 5.6 was exploited to simulate the defined central area resulted from the interference of two opposing SAW. A "Pressure Acoustic, Frequency Domain" is used to analyze the pressure distribution of one-dimensional standing acoustic field[58–62]:

$$\nabla \cdot \left(-\frac{1}{\rho_0} \nabla p\right) = \frac{\omega^2}{\rho_0 c^2}$$

where ω is the angular frequency, $c$ is the sound speed, $\rho_0$ is the density, and $p$ is the pressure. In the permeability computational model, the 70 kDa FITC-dextran solute is represented as a series of discrete beads uniformly distributed on one side of the simulation domain (Fig. 4A green beads). Cooke-Deserno three-bead membrane representation model is implemented to simulate vascular membrane, which offers an elegant balance between computational simplicity and biological realism[63]. The interaction between these entities is defined by employing the Lennard-Jones potential, allowing for the manifestation of Brownian motion among the beads. The use of the Lennard-Jones potential not only provides a mechanism for interaction between the beads but also models the diffusion of these bead-represented solutes through the membrane[64,65]. Within this framework, each lipid is depicted as three discrete beads, representing the hydrophilic head group(red in fig. 4) and the two hydrophobic tail groups(blue in fig. 4), effectively capturing the amphiphilic nature of lipids. These head and tail coarse-grained beads are interconnected via a finite extensible nonlinear elastic bond, thus capturing the essential amphiphilic character of lipids. Additionally, a harmonic angular potential is employed to maintain the straightened configuration of the lipid molecule. The lipid bilayer membrane representative of HUVECs (Human Umbilical Vein Endothelial Cells) and fibroblasts is modeled using two distinct membrane layers, with each layer consisting of three beads. To closely emulate the experimentally observed diffusion rates of membrane permeability, the area per lipid in the membrane was optimized by adjusting the Lennard-Jones parameters. Several foundational assumptions were made to enhance computational efficiency and ensure clarity of the simulation. we conceptualized the lipid bilayer as a seamless, uniform surface, intentionally omitting intricate molecular details reduce computational complexity. Furthermore, the complexities of lipid head-groups and tails are abstracted into coarse-grained beads, encapsulating essential interaction properties. Lastly, the behavior and dynamics of the membrane are primarily driven by a balance between curvature elasticity and thermal fluctuations. The membrane representation and the associated parameters was described in our previous publication[66].

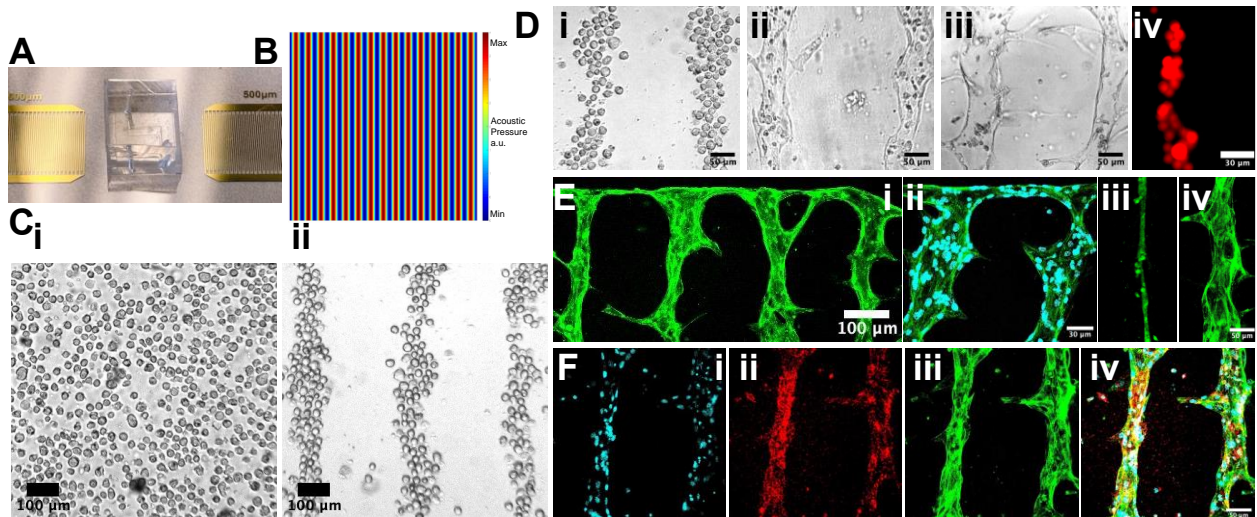

Figure 2. Acoustofluidic engineering of suspended HUVECs into functional vessel-on-a chip. A) The photo of the SSAW device and microfluidic chamber. B) The simulation of the SSAW distribution. C) Before patterning (i), the cells are randomly distributed. After patterning (ii), the cells are aligned into parallel straight-line distribution. D) The separated cells (i) self-assemble at hour 24 (ii) and formed parallel perfusable vessel tubes at hour 48 (iii). 10-um-fluorescent beads can be loaded into the vessel tube (iv). E) The geometry of vessel-on-a-chip(i). The HUVECs are labeled with 488 fluorescence-conjugated-Factin. (ii) The anastomosis structure between the parallel vessel tubes and the side tube perpendicular to them. Cell nuclei is labeled with DAPI. (iii) The vessel tube under static culture condition. (iv) The vessel tube under interstitial flow-stimulated culture condition. F) The vessel tubes are characterized with the immunostaining. Cell nuclei is labeled with DAPI (i). Vessel biomarker is labeled with red CD31 (ii). Cytoskeleton is labeled with green Factin (iii). The overlap image is presented as (iv).

## RESULTS AND DISCUSSION

### Acoustofluidic patterning of the suspended HUVECs into acoustic pressure nodes

Figure. 1 illustrates the methodology workflow of the acoustofluidic vessel-on-a-chip engineering. Once signal of the resonant frequency from the signal generator was amplified and introduced to the acoustofluidic device, two identical but opposite-propagating traveling surface acoustic waves (SAW) were released from the IDT pair, and the SSAW field was constructed on the piezoelectric LiNbO$_3$ wafer surface[67,68]. Thus, a periodic distribution of pressure nodes and antinodes with minimum and maximum pressure amplitudes was presenting respectively (Fig. 2B)[69–71]. With the conduction of the coupling water layer, the acoustic field was introduced into the chamber[72]. As a result, the originally randomly distributed cells was relocated to the nearest pressure nodes, thus forming parallel linear arrays[73]. Here, the half wavelength of the SSAW was designed as 250 μm, and the final patterned cell line spacing was approximately 150 μm (Fig. 2C). Fibrinogen is currently the most suitable hydrogel for culturing vascular tissue *in vitro*, and it was selected as the biological matrix in this experiment[74,75]. The thrombin concentration was reduced to prolong the time required for the gel solidification, until the cells could be fully patterned. In the experiment, it was observed that the fibrinogen gel began to coagulate after being mixed with thrombin for 3 minutes at room temperature, at which time the cells rarely moved. Thus, after 3 minutes, the sample was moved into the incubator for fully solidification. After 3-minute-patterning, the microfluidic chambers were moved into the incubator to fully solidify the gel. In this way, the original cell patterning topography was maintained by the solidified gel could after the acoustic field was withdrawn[76].

**Patterned vessel-on-a-chip formation under interstitial flow-assisted *in vitro* cell culture**

Prior to this project, several papers have been published on the construction of vascular models by acoustic cell patterning[77–79]. For example, Kang *et al.* acoustically patterned endothelial cells in hyaluronic acid to construct vascular tissue for ischemia therapy[80]. The authors transplanted the *in vitro* vascular tissue into a mouse model. After the vascular tissue was anastomosed with the rat's own blood vessels, the vessel persusability was verified *in vivo*. Besides, Zhang *et al.* reported that for *in vitro* vascular models, the key to making vessel tubes perfusable is to provide a stimulus of interstitial flow[81]. The interstitial flow can promote the expression of matrix metalloproteinase-2 protein to enhance the vasculogenic capacity of endothelial cells. Based on these inspiring works, we hypothesized that the perfusable patterned vascular models could be remodeled *in vitro* by delivering the patterned ECs with interstitial flow stimulation.

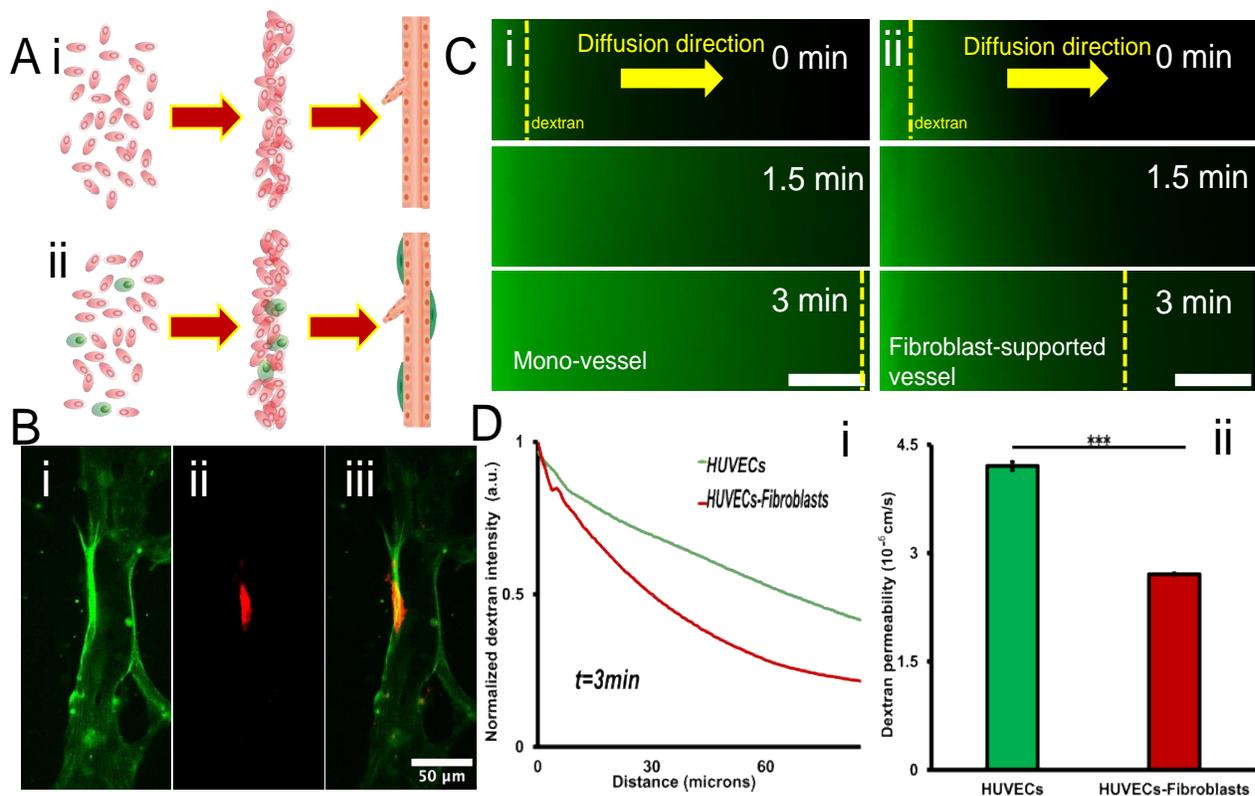

Figure 3. Vascular barrier function characterization with/without supporting fibroblasts. A) Schematics drawing of the vessel formation from the suspended cells. (i) Pure HUVECs situation, and (ii) mixture of UVECs and fibroblasts. B) Coculture of the vessel (i) and the fibroblast (ii). The overlap image (iii) showed the fibroblast is on the outer wall of the vessel. C) Vessel permeability test in mono vessel tube (i) and fibroblast supported vessel tube (ii). Scale bar: 20um. D) (i)70 kDa dextran diffusion profiles at t = 3 min for vessels in various culture conditions. (ii)70 Kda dextran permeability values calculated for vessels in various culture conditions. Error bars show the standard deviation. t-test, *p< 0.05, **p< 0.01, ***p< 0.001, n= 3.

As control, a group of patterned HUVECs and random-seeded HUVECs were cultured in the static culture condition, respectively (Fig. S2). Although HUVECs could also self-assemble into vessel tubes in both groups, these blood vessels are essentially thin and non-perfused. For the experiment group, the patterned HUVECs were applied with hydraulic pressure by loading 50 µl of medium into a 200 µl pipette tip to induce interstitial flow. The patterned ECs began to interconnect in 24 hours (Fig. 2D). The ECs in the lumen not involved in vascularization were washed away in 48 hours (Fig. 2Diii). Parts of ECs migrated to the chamber wall, along which the ECs grew into the horizontal vessel tube to connect all the vertical parallel patterned vessel tubes (Fig. 2Eii). After 48 hours *in vitro* culture, a vascular network was formed, maintaining the shape of the previous acoustic patterning (Fig. 2Ei). The vessel perfusability was demonstrated by loaded the 10 µm red fluorescent beads into the vessel tube (Fig. 2Div). From the cytoskeleton characterization (Fig. 2E), it can be clearly seen that the blood vessels formed under hydraulic pressure stimuli were much wider than the static cultured blood vessels[82] (Fig. 2Eiii&iv, Fig. S4). To verify the HUVEC identity, the patterned vessel tubes were stained with red fluorescence conjugated CD31 antibody (Fig. 2F)[83].

**Vascular barrier function test of the vessel tube with/without the supporting fibroblasts**

In addition to substance transportation, another important function of blood vessels is lateral permeability, which enables the exchange of substances with surrounding tissue, as known as vascular barrier function[84,85]. To evaluate the vascular barrier function in different scenarios, the monocultured vessel was set as control and the fibroblast-supported vessel was set as the experiment group[86]. First, the suspended fibroblasts were mixed into the HUVEC suspension, and went through the acoustofluidic engineering process (Fig. 3A). After 2 days culture, HUVECs could still interconnect and assemble into the straight vessel tubes while the fibroblasts were located on the outer wall of the vessel tube (Fig. S3A). Figure. 3B shows the confocal scanning of the fibroblast-supported vessel structure. The fibroblasts were labeled with red fluorescence conjugated α-smooth muscle actin antibody[87].

Then, the vascular barrier function was quantified by visualizing the diffusion of 70 KDa FITC-dextran solute[88]. The dextran solution was perfused into the parallel patterned vessel tubes through the anastomosis zone (Fig. 2Eii) and permeated into the surrounding matrix. The mono-cultured vessel was leakier compared with the fibroblast-supported vessel (Fig. 3C). The lateral diffusion rate of dextran in the monocultured vessel was almost twice that of the fibroblast-supported vessel (Fig. 3D). For potential interpretation of the interaction between the solute and vascular cell membrane, an *in silico* atomistic model was proposed to visualize how the solute crossing the endothelial cell membrane (Fig. 4).

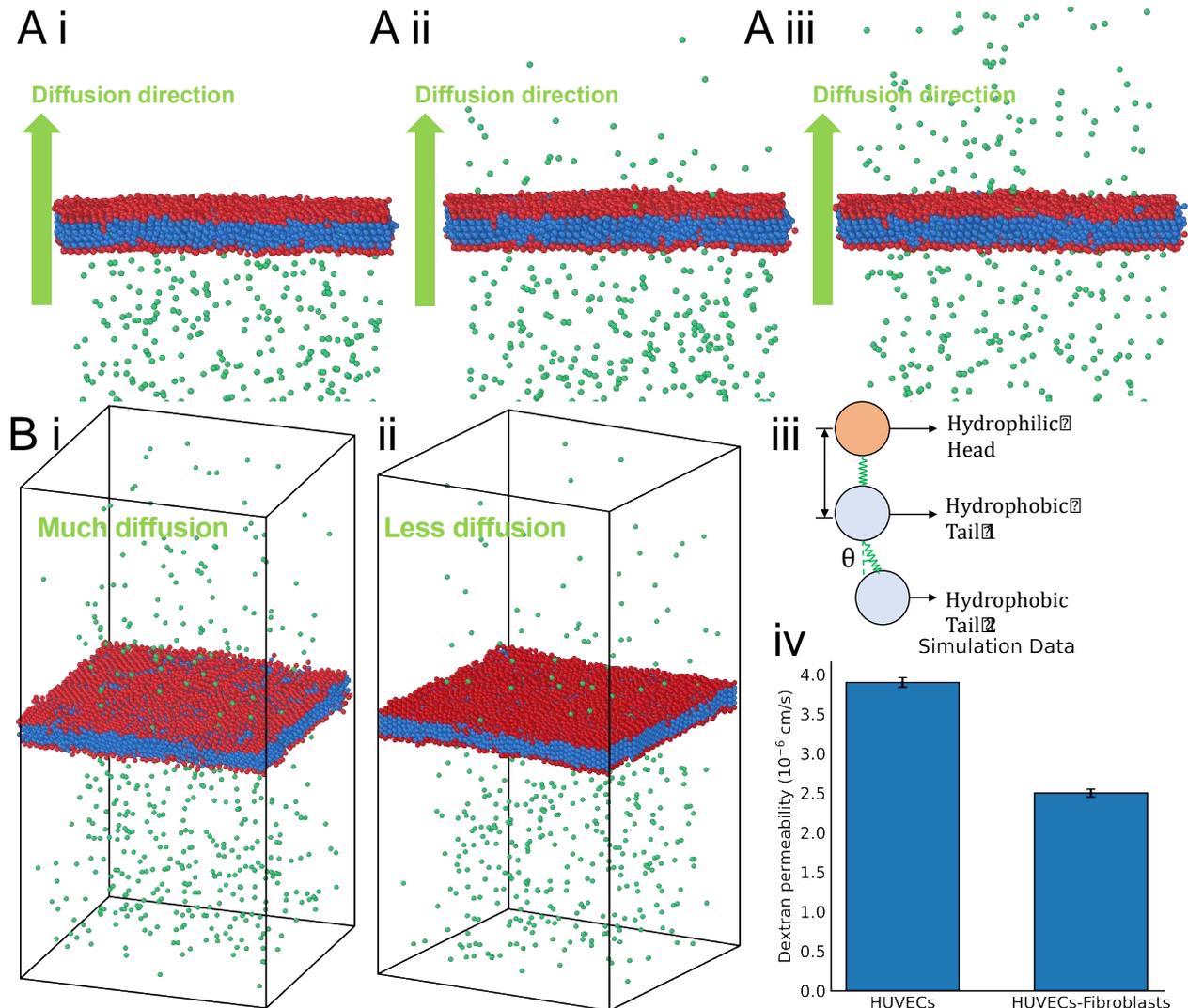

Figure 4. Computational vascular barrier function modeling with/without supporting fibroblasts. A) Modeling of the dextran diffusion process before (i), during (ii) and after (iii) the vessel membrane crossing. B) Modelling the final diffusion result of pure vessel condition within the time unit (i). Modelling the final diffusion result of fibroblast-supported condition within the time unit (ii). (iii) The detailed composition of simulated vascular membrane. (iv) Simulation result of the dextran permeability speed.

**Vascular structure influence compound transport in the vessel-associated matrix**

In addition to demonstrating that surrounding stromal cells can affect blood vessel, it was further shown that the vessel structure can also affect cells in the surrounding environment. Here, a functional assay was designed to show how vascular structure affects compound transport and influences neural activity. Induced pluripotent stem cells (iPSC)-derived neural cells were utilized[89], which express glutamate receptors[50] (Fig. 5). As a control, the iPSC-derived neural cells were resuspended in the pure fibrinogen gel in the microfluidic chamber (Fig. 5Ai). For the experiment group, the neural cells were cocultured with the patterned vessel network (Fig. 5Aii). Glutamate is one of the major excitatory neurotransmitters in the mammalian central nervous system[90]. For the pure hydrogel matrix, solution containing glutamate was loaded into the microfluidic chamber through one inject port. The compound travels through the matrix to reach the suspended cells by passive diffusion[91]. For the vessel-associated matrix, the compound was delivered through the patterned vessel network channel. After 10 minutes of treatment, the samples were washed to remove compound solution. After 20 minutes, the immune-staining for c-Fos was performed. The c-fos gene expression can be activated by a wide range of stimuli and is a reliable marker of neural activity[92]. Its transcription was in rapid manner[93]. In the group in which the compound was delivered by blood vessels, the level of C-fos expression in nerve cells was almost double that of the passive diffusion group (Fig. 5C). In a pure matrix without blood vessels, the substances are transported by passive diffusion, implying slow transport rates, limited transport distances and spatial substance gradients. In a matrix with blood vessel networks, substances are transported directly in vessel tubes, which serve as highway, and reaches the bulk of the matrix more quickly[94].

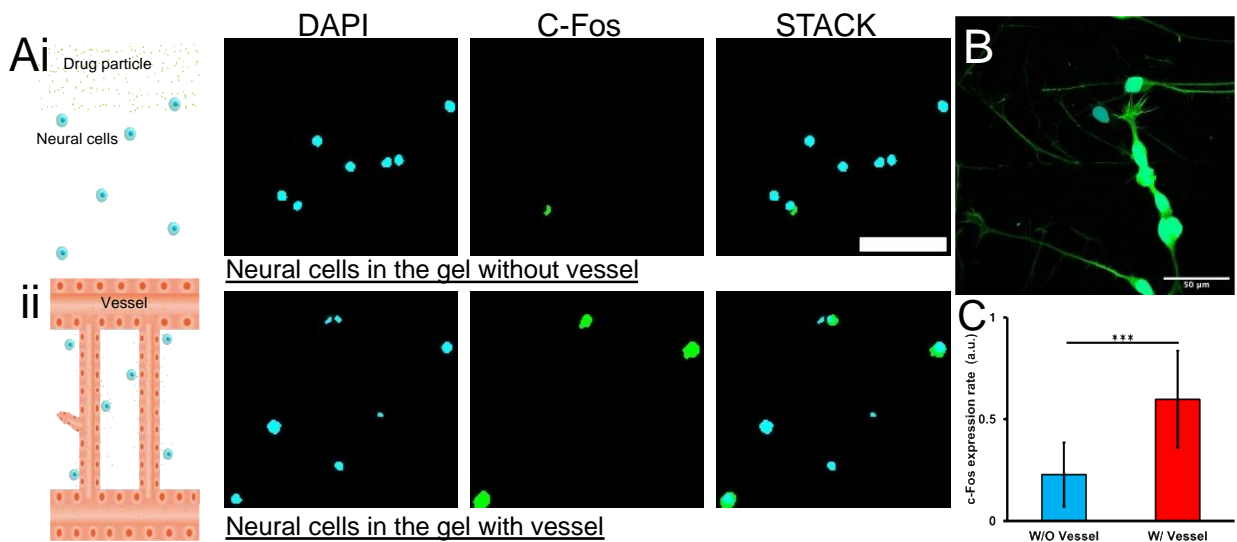

Figure 5. C-Fos expression show reduced how glutamate influence the neural cell without/without vessel transport. A) The neural cells seeded in the hydrogel respond to glutamate stimuli without (i) and with (ii) the vessel transport. Scale bar: 50 microns. B) The neural cells grow in adherence state. C) The normalized c-Fos expression rate with/without vessel transport the compound.

## Conclusion

The reported acoustofluidic engineering of vessel-on-a-chip not only inherited the advantages of previous acoustophoretic publications[80,95], but also incorporated technical points from the literatures about hydraulic pressure activating vessel formation[96,97]. To demonstrate the function of the acoustofluidic engineered vascular structure, the perfusability and permeability of the vessel-on-a-chip was verified. Further, a molecular dynamic computational model was proposed to illustrate how the compound crossing the vascular lipid layer. Lastly, a vessel-neural cells coculture system was established to demonstrate that the vascular structure played the important role in compound delivery process. In general, we reported an acoustofluidic methodology to engineering *in vitro* vessel-on-a-chip model. It is believed that the proposed method can contribute to tissue engineering and regenerative medicine application. In the future, this method has the opportunity to further combine different patterning modes of the sound field (Fig. S5A) to create functional vascular tissues of the specific shapes[98,99].

## ASSOCIATED CONTENT

### Supplementary Materials

Supplementary Figure: Various HUVEC self-assembly behavior under different in vitro culture conditions.

## AUTHOR INFORMATION

### Corresponding Author

Yaling Liu- Department of Mechanical Engineering & Mechanics; Department of Bioengineering; Lehigh University; 558 Packard Lab,19 Memorial Drive West, Bethlehem, PA 18015. Email: yal310@lehigh.edu.

**Notes**

The authors declare no competing financial interest.

**Acknowledgements**

This work was supported by National Institute of Health grant R01HL131750, R21EB033102, National Science Foundation grant CBET 2039310, OAC 2215789, Pennsylvania Department of Health Commonwealth Universal Research Enhancement Program (CURE), and Pennsylvania Infrastructure Technology Alliance (PITA).

**References**

(1) Pugsley, M. K.; Tabrizchi, R. The Vascular System. An Overview of Structure and Function. *J. Pharmacol. Toxicol. Methods* **2000**, *44* (2), 333–340.

(2) Banks, W. A. From Blood-Brain Barrier to Blood-Brain Interface: New Opportunities for CNS Drug Delivery. *Nat. Rev. Drug Discov.* **2016**, *15* (4), 275–292.

(3) Carmeliet, P.; Conway, E. M. Growing Better Blood Vessels. *Nat. Biotechnol.* **2001**, *19* (11), 1019–1020.

(4) Hellmut G, A.; Gou Young, K. Organotypic Vasculature: From Descriptive Heterogeneity to Functional Pathophysiology. *Science* **2017**, *357* (6353), eaal2379.

(5) Potente, M.; Mäkinen, T. Vascular Heterogeneity and Specialization in Development and Disease. *Nat. Rev. Mol. Cell Biol.* **2017**, *18* (8), 477–494.

(6) Lorente, S.; Hautefeuille, M.; Sanchez-Cedillo, A. The Liver, a Functionalized Vascular Structure. *Sci. Rep.* **2020**, *10* (1), 16194.

(7) Jain, R. K.; Au, P.; Tam, J.; Duda, D. G.; Fukumura, D. Engineering Vascularized Tissue. *Nat. Biotechnol.* **2005**, *23* (7), 821–823.

(8) Kwak, T. J.; Lee, E. In Vitro Modeling of Solid Tumor Interactions with Perfused Blood Vessels. *Sci. Rep.* **2020**, *10* (1), 20142.

(9) Haase, K.; Piatti, F.; Marcano, M.; Shin, Y.; Visone, R.; Redaelli, A.; Rasponi, M.; Kamm, R. D. Physiologic Flow-Conditioning Limits Vascular Dysfunction in Engineered Human Capillaries. *Biomaterials* **2022**, *280* (121248), 121248.

(10) Yu, J.; Lee, S.; Song, J.; Lee, S.-R.; Kim, S.; Choi, H.; Kang, H.; Hwang, Y.; Hong, Y.-K.; Jeon, N. L. Perfusable Micro-Vascularized 3D Tissue Array for High-Throughput Vascular Phenotypic Screening. *Nano Converg.* **2022**, *9* (1), 16.

(11) Paek, J.; Park, S. E.; Lu, Q.; Park, K.-T.; Cho, M.; Oh, J. M.; Kwon, K. W.; Yi, Y.-S.; Song, J. W.; Edelstein, H. I.; Ishibashi, J.; Yang, W.; Myerson, J. W.; Kiseleva, R. Y.; Aprelev, P.;

Hood, E. D.; Stambolian, D.; Seale, P.; Muzykantov, V. R.; Huh, D. Microphysiological Engineering of Self-Assembled and Perfusable Microvascular Beds for the Production of Vascularized Three-Dimensional Human Microtissues. *ACS Nano* **2019**, *13* (7), 7627–7643.

(12) Moya, M. L.; Hsu, Y.-H.; Lee, A. P.; Hughes, C. C. W.; George, S. C. In Vitro Perfused Human Capillary Networks. *Tissue Eng. Part C Methods* **2013**, *19* (9), 730–737.

(13) Zhou, Y.; Wu, Y.; Paul, R.; Qin, X.; Liu, Y. Hierarchical Vessel Network-Supported Tumor Model-on-a-Chip Constructed by Induced Spontaneous Anastomosis. *ACS Appl. Mater. Interfaces* **2023**. https://doi.org/10.1021/acsami.2c19453.

(14) Campisi, M.; Shin, Y.; Osaki, T.; Hajal, C.; Chiono, V.; Kamm, R. D. 3D Self-Organized Microvascular Model of the Human Blood-Brain Barrier with Endothelial Cells, Pericytes and Astrocytes. *Biomaterials* **2018**, *180*, 117–129.

(15) *De LumeNEXT: A Practical Method to Pattern Luminal Structures in ECM Gels*.

(16) de Graaf, M. N. S.; Vivas, A.; Kasi, D. G.; van den Hil, F. E.; van den Berg, A.; van der Meer, A. D.; Mummery, C. L.; Orlova, V. V. Multiplexed Fluidic Circuit Board for Controlled Perfusion of 3D Blood Vessels-on-a-Chip. *Lab Chip* **2022**, *23* (1), 168–181.

(17) Wu, Y.; Zhou, Y.; Paul, R.; Qin, X.; Islam, K.; Liu, Y. Adaptable Microfluidic Vessel-on-a-Chip Platform for Investigating Tumor Metastatic Transport in Bloodstream. *Anal. Chem.* **2022**. https://doi.org/10.1021/acs.analchem.2c02556.

(18) Zhao, N.; Guo, Z.; Kulkarni, S.; Norman, D.; Zhang, S.; Chung, T. D.; Nerenberg, R. F.; Linville, R.; Searson, P. Engineering the Human Blood-Brain Barrier at the Capillary Scale Using a Double-Templating Technique. *Adv. Funct. Mater.* **2022**, *32* (30), 2110289.

(19) Paul, R.; Zhao, Y.; Coster, D.; Qin, X.; Islam, K.; Wu, Y.; Liu, Y. Rapid Prototyping of High-Resolution Large Format Microfluidic Device through Maskless Image Guided in-Situ Photopolymerization. *Nat. Commun.* **2023**, *14* (1), 4520.

(20) Rayner, S. G.; Howard, C. C.; Mandrycky, C. J.; Stamenkovic, S.; Himmelfarb, J.; Shih, A. Y.; Zheng, Y. Multiphoton-Guided Creation of Complex Organ-Specific Microvasculature. *Adv. Healthc. Mater.* **2021**, *10* (10), e2100031.

(21) Arakawa, C.; Gunnarsson, C.; Howard, C.; Bernabeu, M.; Phong, K.; Yang, E.; DeForest, C. A.; Smith, J. D.; Zheng, Y. Biophysical and Biomolecular Interactions of Malaria-Infected Erythrocytes in Engineered Human Capillaries. *Sci. Adv.* **2020**, *6* (3), eaay7243.

(22) Arakawa, C. K.; Badeau, B. A.; Zheng, Y.; DeForest, C. A. Multicellular Vascularized Engineered Tissues through User-Programmable Biomaterial Photodegradation. *Adv. Mater.* **2017**, *29* (37). https://doi.org/10.1002/adma.201703156.

(23) Ozcelik, A.; Rufo, J.; Guo, F.; Gu, Y.; Li, P.; Lata, J.; Huang, T. J. Acoustic Tweezers for the Life Sciences. *Nat. Methods* **2018**, *15* (12), 1021–1028.

(24) Yang, S.; Rufo, J.; Zhong, R.; Rich, J.; Wang, Z.; Lee, L. P.; Huang, T. J. Acoustic Tweezers for High-Throughput Single-Cell Analysis. *Nat. Protoc.* **2023**, *18* (8), 2441–2458.

(25) Li, P.; Mao, Z.; Peng, Z.; Zhou, L.; Chen, Y.; Huang, P.-H.; Truica, C. I.; Drabick, J. J.; El-Deiry, W. S.; Dao, M.; Suresh, S.; Huang, T. J. Acoustic Separation of Circulating Tumor Cells. *Proc. Natl. Acad. Sci. U. S. A.* **2015**, *112* (16), 4970–4975.

(26) Wu, M.; Ouyang, Y.; Wang, Z.; Zhang, R.; Huang, P.-H.; Chen, C.; Li, H.; Li, P.; Quinn, D.; Dao, M.; Suresh, S.; Sadovsky, Y.; Huang, T. J. Isolation of Exosomes from Whole Blood by Integrating Acoustics and Microfluidics. *Proc. Natl. Acad. Sci. U. S. A.* **2017**, *114* (40), 10584–10589.

(27) Gu, Y.; Chen, C.; Mao, Z.; Bachman, H.; Becker, R.; Rufo, J.; Wang, Z.; Zhang, P.; Mai, J.; Yang, S.; Zhang, J.; Zhao, S.; Ouyang, Y.; Wong, D. T. W.; Sadovsky, Y.; Huang, T. J. Acoustofluidic Centrifuge for Nanoparticle Enrichment and Separation. *Sci Adv* **2021**, *7* (1). https://doi.org/10.1126/sciadv.abc0467.

(28) Hao, N.; Pei, Z.; Liu, P.; Bachman, H.; Naquin, T. D.; Zhang, P.; Zhang, J.; Shen, L.; Yang, S.; Yang, K.; Zhao, S.; Huang, T. J. Acoustofluidics-Assisted Fluorescence-SERS Bimodal Biosensors. *Small* **2020**, *16* (48), e2005179.

(29) Wang, Z.; Li, F.; Rufo, J.; Chen, C.; Yang, S.; Li, L.; Zhang, J.; Cheng, J.; Kim, Y.; Wu, M.; Abemayor, E.; Tu, M.; Chia, D.; Spruce, R.; Batis, N.; Mehanna, H.; Wong, D. T. W.; Huang, T. J. Acoustofluidic Salivary Exosome Isolation: A Liquid Biopsy Compatible Approach for Human Papillomavirus-Associated Oropharyngeal Cancer Detection. *J. Mol. Diagn.* **2020**, *22* (1), 50–59.

(30) Ma, Z.; Zhou, Y.; Collins, D. J.; Ai, Y. Fluorescence Activated Cell Sorting via a Focused Traveling Surface Acoustic Beam. *Lab Chip* **2017**, *17* (18), 3176–3185.

(31) Xu, J.; Cai, H.; Wu, Z.; Li, X.; Tian, C.; Ao, Z.; Niu, V. C.; Xiao, X.; Jiang, L.; Khodoun, M.; Rothenberg, M.; Mackie, K.; Chen, J.; Lee, L. P.; Guo, F. Acoustic Metamaterials-Driven Transdermal Drug Delivery for Rapid and on-Demand Management of Acute Disease. *Nat. Commun.* **2023**, *14* (1), 869.

(32) Ramesan, S.; Rezk, A. R.; Dekiwadia, C.; Cortez-Jugo, C.; Yeo, L. Y. Acoustically-Mediated Intracellular Delivery. *Nanoscale* **2018**, *10* (27), 13165–13178.

(33) Banerjee, T.; Gosai, A.; Yousefi, N.; Garibay, O. O.; Seal, S.; Balasubramanian, G. Examining Sialic Acid Derivatives as Potential Inhibitors of SARS-CoV-2 Spike Protein Receptor Binding Domain. *J. Biomol. Struct. Dyn.* **2023**, 1–17.

(34) Yousefi, N.; Yazdani-Jahromi, M.; Tayebi, A.; Kolanthai, E.; Neal, C. J.; Banerjee, T.; Gosai, A.; Balasubramanian, G.; Seal, S.; Ozmen Garibay, O. BindingSite-AugmentedDTA: Enabling a next-Generation Pipeline for Interpretable Prediction Models in Drug Repurposing. *Brief. Bioinform.* **2023**, *24* (3). https://doi.org/10.1093/bib/bbad136.

(35) Wu, Z.; Ao, Z.; Cai, H.; Li, X.; Chen, B.; Tu, H.; Wang, Y.; Lu, R. O.; Gu, M.; Cheng, L.; Lu, X.; Guo, F. Acoustofluidic Assembly of Primary Tumor-Derived Organotypic Cell Clusters for Rapid Evaluation of Cancer Immunotherapy. *J. Nanobiotechnology* **2023**, *21* (1), 40.

(36) Ao, Z.; Wu, Z.; Cai, H.; Hu, L.; Li, X.; Kaurich, C.; Chang, J.; Gu, M.; Cheng, L.; Lu, X.; Guo, F. Rapid Profiling of Tumor-Immune Interaction Using Acoustically Assembled Patient-Derived Cell Clusters. *Adv. Sci. (Weinh.)* **2022**, *9* (22), e2201478.

(37) Armstrong, J. P. K.; Pchelintseva, E.; Treumuth, S.; Campanella, C.; Meinert, C.; Klein, T. J.; Hutmacher, D. W.; Drinkwater, B. W.; Stevens, M. M. Tissue Engineering Cartilage with Deep Zone Cytoarchitecture by High-Resolution Acoustic Cell Patterning. *Adv. Healthc. Mater.* **2022**, e2200481.

(38) Armstrong, J. P. K.; Puetzer, J. L.; Serio, A.; Guex, A. G.; Kapnisi, M.; Breant, A.; Zong, Y.; Assal, V.; Skaalure, S. C.; King, O.; Murty, T.; Meinert, C.; Franklin, A. C.; Bassindale, P. G.; Nichols, M. K.; Terracciano, C. M.; Hutmacher, D. W.; Drinkwater, B. W.; Klein, T. J.; Perriman, A. W.; Stevens, M. M. Engineering Anisotropic Muscle Tissue Using Acoustic Cell Patterning. *Adv. Mater.* **2018**, *30* (43), e1802649.

(39) Bouyer, C.; Chen, P.; Güven, S.; Demirtaş, T. T.; Nieland, T. J. F.; Padilla, F.; Demirci, U. A Bio-Acoustic Levitational (BAL) Assembly Method for Engineering of Multilayered, 3D Brain-like Constructs, Using Human Embryonic Stem Cell Derived Neuro-Progenitors. *Adv. Mater.* **2016**, *28* (1), 161–167.

(40) Chen, B.; Wu, Z.; Wu, Y.; Chen, Y.; Zheng, L. Controllable Fusion of Multicellular Spheroids Using Acoustofluidics. *Microfluid. Nanofluidics* **2023**, *27* (7). https://doi.org/10.1007/s10404-023-02660-5.

(41) Link, A.; Franke, T. Acoustic Erythrocytometer for Mechanically Probing Cell Viscoelasticity. *Lab Chip* **2020**, *20* (11), 1991–1998.

(42) Cai, H.; Ao, Z.; Wu, Z.; Nunez, A.; Jiang, L.; Carpenter, R. L.; Nephew, K. P.; Guo, F. Profiling Cell-Matrix Adhesion Using Digitalized Acoustic Streaming. *Anal. Chem.* **2020**, *92* (2), 2283–2290.

(43) Gai, J.; Nosrati, R.; Neild, A. High DNA Integrity Sperm Selection Using Surface Acoustic Waves. *Lab Chip* **2020**, *20* (22), 4262–4272.

(44) Gai, J.; Devendran, C.; Neild, A.; Nosrati, R. Surface Acoustic Wave-Driven Pumpless Flow for Sperm Rheotaxis Analysis. *Lab Chip* **2022**, *22* (22), 4409–4417.

(45) Liao, Q.-Q.; Zhao, S.-K.; Cai, B.; He, R.-X.; Rao, L.; Wu, Y.; Guo, S.-S.; Liu, Q.-Y.; Liu, W.; Zhao, X.-Z. Biocompatible Fabrication of Cell-Laden Calcium Alginate Microbeads Using Microfluidic Double Flow-Focusing Device. *Sens. Actuators A Phys.* **2018**, *279*, 313–320.

(46) Guo, F.; Xie, Y.; Li, S.; Lata, J.; Ren, L.; Mao, Z.; Ren, B.; Wu, M.; Ozcelik, A.; Huang, T. J. Reusable Acoustic Tweezers for Disposable Devices. *Lab Chip* **2015**, *15* (24), 4517–4523.

(47) Wu, Y.; Zhao, Y.; Zhou, Y.; Islam, K.; Liu, Y. Microfluidic Droplet-Assisted Fabrication of Vessel-Supported Tumors for Preclinical Drug Discovery. *ACS Appl. Mater. Interfaces* **2023**, *15* (12), 15152–15161.

(48) Hajal, C.; Offeddu, G. S.; Shin, Y.; Zhang, S.; Morozova, O.; Hickman, D.; Knutson, C. G.; Kamm, R. D. Engineered Human Blood-Brain Barrier Microfluidic Model for Vascular Permeability Analyses. *Nat. Protoc.* **2022**, *17* (1), 95–128.

(49) Pak, C.; Pak, C.; Grieder, S.; Yang, N.; Zhang, Y.; Wernig, M.; Sudhof, T. Rapid Generation of Functional and Homogeneous Excitatory Human Forebrain Neurons Using Neurogenin-2 (Ngn2). *Protoc. Exch.* **2018**. https://doi.org/10.1038/protex.2018.082.

(50) Zhang, Y.; Pak, C.; Han, Y.; Ahlenius, H.; Zhang, Z.; Chanda, S.; Marro, S.; Patzke, C.; Acuna, C.; Covy, J.; Xu, W.; Yang, N.; Danko, T.; Chen, L.; Wernig, M.; Südhof, T. C. Rapid Single-Step Induction of Functional Neurons from Human Pluripotent Stem Cells. *Neuron* **2013**, *78* (5), 785–798.

(51) Yeon, J. H.; Ryu, H. R.; Chung, M.; Hu, Q. P.; Jeon, N. L. In Vitro Formation and Characterization of a Perfusable Three-Dimensional Tubular Capillary Network in Microfluidic Devices. *Lab Chip* **2012**, *12* (16), 2815–2822.

(52) Chen, B.; Wu, Y.; Ao, Z.; Cai, H.; Nunez, A.; Liu, Y.; Foley, J.; Nephew, K.; Lu, X.; Guo, F. High-Throughput Acoustofluidic Fabrication of Tumor Spheroids. *Lab Chip* **2019**, *19* (10), 1755–1763.

(53) Lata, J. P.; Guo, F.; Guo, J.; Huang, P.-H.; Yang, J.; Huang, T. J. Surface Acoustic Waves Grant Superior Spatial Control of Cells Embedded in Hydrogel Fibers. *Adv. Mater.* **2016**, *28* (39), 8632–8638.

(54) Wu, Z.; Chen, B.; Wu, Y.; Xia, Y.; Chen, H.; Gong, Z.; Hu, H.; Ding, Z.; Guo, S. Scaffold-Free Generation of Heterotypic Cell Spheroids Using Acoustofluidics. *Lab Chip* **2021**. https://doi.org/10.1039/d1lc00496d.

(55) Katche, C.; Bekinschtein, P.; Slipczuk, L.; Goldin, A.; Izquierdo, I. A.; Cammarota, M.; Medina, J. H. Delayed Wave of C-Fos Expression in the Dorsal Hippocampus Involved

Specifically in Persistence of Long-Term Memory Storage. *Proc. Natl. Acad. Sci. U. S. A.* **2010**, *107* (1), 349–354.

(56) Virumbrales-Muñoz, M.; Chen, J.; Ayuso, J.; Lee, M.; Abel, E. J.; Beebe, D. J. Organotypic Primary Blood Vessel Models of Clear Cell Renal Cell Carcinoma for Single-Patient Clinical Trials. *Lab Chip* **2020**, *20* (23), 4420–4432.

(57) Ingram, P. N.; Hind, L. E.; Jiminez-Torres, J. A.; Huttenlocher, A.; Beebe, D. J. An Accessible Organotypic Microvessel Model Using IPSC-Derived Endothelium. *Adv. Healthc. Mater.* **2018**, *7* (2), 1700497.

(58) Ni, Z.; Yin, C.; Xu, G.; Xie, L.; Huang, J.; Liu, S.; Tu, J.; Guo, X.; Zhang, D. Modelling of SAW-PDMS Acoustofluidics: Physical Fields and Particle Motions Influenced by Different Descriptions of the PDMS Domain. *Lab Chip* **2019**, *19* (16), 2728–2740.

(59) Cai, H.; Ao, Z.; Hu, L.; Moon, Y.; Wu, Z.; Lu, H.-C.; Kim, J.; Guo, F. Acoustofluidic Assembly of 3D Neurospheroids to Model Alzheimer's Disease. *Analyst* **2020**, *145* (19), 6243–6253.

(60) Pan, H.; Mei, D.; Xu, C.; Han, S.; Wang, Y. Bisymmetric Coherent Acoustic Tweezers Based on Modulation of Surface Acoustic Waves for Dynamic and Reconfigurable Cluster Manipulation of Particles and Cells. *Lab Chip* **2023**, *23* (2), 215–228.

(61) Nama, N.; Barnkob, R.; Mao, Z.; Kähler, C. J.; Costanzo, F.; Huang, T. J. Numerical Study of Acoustophoretic Motion of Particles in a PDMS Microchannel Driven by Surface Acoustic Waves. *Lab Chip* **2015**, *15* (12), 2700–2709.

(62) Hsu, J.-C.; Chao, C.-L. Acoustophoretic Patterning of Microparticles in a Microfluidic Chamber Driven by Standing Lamb Waves. *Appl. Phys. Lett.* **2021**, *119* (10), 103504.

(63) Cooke, I. R.; Kremer, K.; Deserno, M. Tunable Generic Model for Fluid Bilayer Membranes. *Phys. Rev. E Stat. Nonlin. Soft Matter Phys.* **2005**, *72* (1 Pt 1), 011506.

(64) Razizadeh, M.; Nikfar, M.; Paul, R.; Liu, Y. Coarse-Grained Modeling of Pore Dynamics on the Red Blood Cell Membrane under Large Deformations. *Biophys. J.* **2020**, *119* (3), 471–482.

(65) Jorgensen, C.; Ulmschneider, M. B.; Searson, P. C. Atomistic Model of Solute Transport across the Blood-Brain Barrier. *ACS Omega* **2022**, *7* (1), 1100–1112.

(66) Islam, K.; Razizadeh, M.; Liu, Y. Coarse-Grained Molecular Simulation of Extracellular Vesicle Squeezing for Drug Loading. *Phys. Chem. Chem. Phys.* **2023**, *25* (17), 12308–12321.

(67) Guo, F.; Li, P.; French, J. B.; Mao, Z.; Zhao, H.; Li, S.; Nama, N.; Fick, J. R.; Benkovic, S. J.; Huang, T. J. Controlling Cell-Cell Interactions Using Surface Acoustic Waves. *Proc. Natl. Acad. Sci. U. S. A.* **2015**, *112* (1), 43–48.

(68) Guo, F.; Mao, Z.; Chen, Y.; Xie, Z.; Lata, J. P.; Li, P.; Ren, L.; Liu, J.; Yang, J.; Dao, M.; Suresh, S.; Huang, T. J. Three-Dimensional Manipulation of Single Cells Using Surface Acoustic Waves. *Proc. Natl. Acad. Sci. U. S. A.* **2016**, *113* (6), 1522–1527.

(69) Ding, X.; Shi, J.; Lin, S.-C. S.; Yazdi, S.; Kiraly, B.; Huang, T. J. Tunable Patterning of Microparticles and Cells Using Standing Surface Acoustic Waves. *Lab Chip* **2012**, *12* (14), 2491–2497.

(70) Mao, Z.; Xie, Y.; Guo, F.; Ren, L.; Huang, P.-H.; Chen, Y.; Rufo, J.; Costanzo, F.; Huang, T. J. Experimental and Numerical Studies on Standing Surface Acoustic Wave Microfluidics. *Lab Chip* **2016**, *16* (3), 515–524.

(71) Guo, F.; Zhou, W.; Li, P.; Mao, Z.; Yennawar, N. H.; French, J. B.; Huang, T. J. Precise Manipulation and Patterning of Protein Crystals for Macromolecular Crystallography Using Surface Acoustic Waves. *Small* **2015**, *11* (23), 2733–2737.

(72) Wu, Y.; Ao, Z.; Chen, B.; Muhsen, M.; Bondesson, M.; Lu, X.; Guo, F. Acoustic Assembly of Cell Spheroids in Disposable Capillaries. *Nanotechnology* **2018**, *29* (50), 504006.

(73) Li, S.; Guo, F.; Chen, Y.; Ding, X.; Li, P.; Wang, L.; Cameron, C. E.; Huang, T. J. Standing Surface Acoustic Wave Based Cell Coculture. *Anal. Chem.* **2014**, *86* (19), 9853–9859.

(74) Chen, M. B.; Whisler, J. A.; Fröse, J.; Yu, C.; Shin, Y.; Kamm, R. D. On-Chip Human Microvasculature Assay for Visualization and Quantification of Tumor Cell Extravasation Dynamics. *Nat. Protoc.* **2017**, *12* (5), 865–880.

(75) Wan, Z.; Zhong, A. X.; Zhang, S.; Pavlou, G.; Coughlin, M. F.; Shelton, S. E.; Nguyen, H. T.; Lorch, J. H.; Barbie, D. A.; Kamm, R. D. A Robust Method for Perfusable Microvascular Network Formation in Vitro. *Small Methods* **2022**, *6* (6), e2200143.

(76) Tian, Z.; Wang, Z.; Zhang, P.; Naquin, T. D.; Mai, J.; Wu, Y.; Yang, S.; Gu, Y.; Bachman, H.; Liang, Y.; Yu, Z.; Huang, T. J. Generating Multifunctional Acoustic Tweezers in Petri Dishes for Contactless, Precise Manipulation of Bioparticles. *Sci. Adv.* **2020**, *6* (37). https://doi.org/10.1126/sciadv.abb0494.

(77) Petta, D.; Basoli, V.; Pellicciotta, D.; Tognato, R.; Barcik, J.; Arrigoni, C.; Bella, E. D.; Armiento, A. R.; Candrian, C.; Richards, R. G.; Alini, M.; Moretti, M.; Eglin, D.; Serra, T. Sound-Induced Morphogenesis of Multicellular Systems for Rapid Orchestration of Vascular Networks. *Biofabrication* **2020**, *13* (1), 015004.

(78) Di Marzio, N.; Ananthanarayanan, P.; Guex, A. G.; Alini, M.; Riganti, C.; Serra, T. Sound-Based Assembly of a Microcapillary Network in a Saturn-like Tumor Model for Drug Testing. *Mater. Today Bio* **2022**, *16* (100357), 100357.

(79) Garvin, K. A.; Dalecki, D.; Yousefhussien, M.; Helguera, M.; Hocking, D. C. Spatial Patterning of Endothelial Cells and Vascular Network Formation Using Ultrasound Standing Wave Fields. *J. Acoust. Soc. Am.* **2013**, *134* (2), 1483–1490.

(80) Kang, B.; Shin, J.; Park, H.-J.; Rhyou, C.; Kang, D.; Lee, S.-J.; Yoon, Y.-S.; Cho, S.-W.; Lee, H. High-Resolution Acoustophoretic 3D Cell Patterning to Construct Functional Collateral Cylindroids for Ischemia Therapy. *Nat. Commun.* **2018**, *9* (1). https://doi.org/10.1038/s41467-018-07823-5.

(81) Zhang, S.; Wan, Z.; Pavlou, G.; Zhong, A. X.; Xu, L.; Kamm, R. D. Interstitial Flow Promotes the Formation of Functional Microvascular Networks in Vitro through Upregulation of Matrix Metalloproteinase‐2. *Adv. Funct. Mater.* **2022**, 2206767.

(82) Winkelman, M. A.; Kim, D. Y.; Kakarla, S.; Grath, A.; Silvia, N.; Dai, G. Interstitial Flow Enhances the Formation, Connectivity, and Function of 3D Brain Microvascular Networks Generated within a Microfluidic Device. *Lab Chip* **2021**, *22* (1), 170–192.

(83) Vila Cuenca, M.; Cochrane, A.; van den Hil, F. E.; de Vries, A. A. F.; Lesnik Oberstein, S. A. J.; Mummery, C. L.; Orlova, V. V. Engineered 3D Vessel-on-Chip Using HiPSC-Derived Endothelial- and Vascular Smooth Muscle Cells. *Stem Cell Reports* **2021**, *16* (9), 2159–2168.

(84) Offeddu, G. S.; Haase, K.; Gillrie, M. R.; Li, R.; Morozova, O.; Hickman, D.; Knutson, C. G.; Kamm, R. D. An On-Chip Model of Protein Paracellular and Transcellular Permeability in the Microcirculation. *Biomaterials* **2019**, *212*, 115–125.

(85) Ho, Y. T.; Adriani, G.; Beyer, S.; Nhan, P.-T.; Kamm, R. D.; Kah, J. C. Y. A Facile Method to Probe the Vascular Permeability of Nanoparticles in Nanomedicine Applications. *Sci. Rep.* **2017**, *7* (1). https://doi.org/10.1038/s41598-017-00750-3.

(86) Margolis, E. A.; Cleveland, D. S.; Kong, Y. P.; Beamish, J. A.; Wang, W. Y.; Baker, B. M.; Putnam, A. J. Stromal Cell Identity Modulates Vascular Morphogenesis in a Microvasculature-on-a-Chip Platform. *Lab Chip* **2021**, *21* (6), 1150–1163.

(87) Pal, D.; Ghatak, S.; Singh, K.; Abouhashem, A. S.; Kumar, M.; El Masry, M. S.; Mohanty, S. K.; Palakurti, R.; Rustagi, Y.; Tabasum, S.; Khona, D. K.; Khanna, S.; Kacar, S.; Srivastava, R.; Bhasme, P.; Verma, S. S.; Hernandez, E.; Sharma, A.; Reese, D.; Verma, P.; Ghosh, N.; Gorain, M.; Wan, J.; Liu, S.; Liu, Y.; Castro, N. H.; Gnyawali, S. C.; Lawrence, W.; Moore, J.; Perez, D. G.; Roy, S.; Yoder, M. C.; Sen, C. K. Identification of a Physiologic Vasculogenic Fibroblast State to Achieve Tissue Repair. *Nat. Commun.* **2023**, *14* (1), 1129.

(88) Zheng, Y.; Chen, J.; Craven, M.; Choi, N. W.; Totorica, S.; Diaz-Santana, A.; Kermani, P.; Hempstead, B.; Fischbach-Teschl, C.; López, J. A.; Stroock, A. D. In Vitro Microvessels for the Study of Angiogenesis and Thrombosis. *Proc. Natl. Acad. Sci. U. S. A.* **2012**, *109* (24), 9342–9347.

(89) Hasan, M. F.; Berdichevsky, Y. Neuron and Astrocyte Aggregation and Sorting in Three-Dimensional Neuronal Constructs. *Commun. Biol.* **2021**, *4* (1), 587.

(90) Zhou, Y.; Danbolt, N. C. Glutamate as a Neurotransmitter in the Healthy Brain. *J. Neural Transm. (Vienna)* **2014**, *121* (8), 799–817.

(91) Adriani, G.; Ma, D.; Pavesi, A.; Kamm, R. D.; Goh, E. L. K. A 3D Neurovascular Microfluidic Model Consisting of Neurons, Astrocytes and Cerebral Endothelial Cells as a Blood–Brain Barrier. *Lab Chip* **2017**, *17* (3), 448–459.

(92) Joo, J.-Y.; Schaukowitch, K.; Farbiak, L.; Kilaru, G.; Kim, T.-K. Stimulus-Specific Combinatorial Functionality of Neuronal c-Fos Enhancers. *Nat. Neurosci.* **2016**, *19* (1), 75–83.

(93) Bullitt, E. Expression of C-Fos-like Protein as a Marker for Neuronal Activity Following Noxious Stimulation in the Rat. *J. Comp. Neurol.* **1990**, *296* (4), 517–530.

(94) Rademakers, T.; Horvath, J. M.; van Blitterswijk, C. A.; LaPointe, V. L. S. Oxygen and Nutrient Delivery in Tissue Engineering: Approaches to Graft Vascularization. *J. Tissue Eng. Regen. Med.* **2019**, *13* (10), 1815–1829.

(95) Hu, X.; Zheng, J.; Hu, Q.; Liang, L.; Yang, D.; Cheng, Y.; Li, S.-S.; Chen, L.-J.; Yang, Y. Smart Acoustic 3D Cell Construct Assembly with High-Resolution. *Biofabrication* **2022**, *14* (4), 045003.

(96) Wan, H.-Y.; Chen, J. C. H.; Xiao, Q.; Wong, C. W.; Yang, B.; Cao, B.; Tuan, R. S.; Nilsson, S. K.; Ho, Y.-P.; Raghunath, M.; Kamm, R. D.; Blocki, A. Stabilization and Improved Functionality of Three-Dimensional Perfusable Microvascular Networks in Microfluidic Devices under Macromolecular Crowding. *Biomater. Res.* **2023**, *27* (1), 32.

(97) Wan, Z.; Floryan, M. A.; Coughlin, M. F.; Zhang, S.; Zhong, A. X.; Shelton, S. E.; Wang, X.; Xu, C.; Barbie, D. A.; Kamm, R. D. New Strategy for Promoting Vascularization in Tumor Spheroids in a Microfluidic Assay. *Adv. Healthc. Mater.* **2022**, e2201784.

(98) Pan, H.; Mei, D.; Xu, C.; Li, X.; Wang, Y. Acoustic Tweezers Using Bisymmetric Coherent Surface Acoustic Waves for Dynamic and Reconfigurable Manipulation of Particle Multimers. *J. Colloid Interface Sci.* **2023**, *643*, 115–123.

(99) Hu, X.; Zhu, J.; Zuo, Y.; Yang, D.; Zhang, J.; Cheng, Y.; Yang, Y. Versatile Biomimetic Array Assembly by Phase Modulation of Coherent Acoustic Waves. *Lab Chip* **2020**, *20* (19), 3515–3523.